\documentclass{PoS}

\title{Heavy MSSM Higgs Interpretation of the\\ LHC Run\,I Data}

\usepackage{amsmath,amssymb,amsfonts,color,graphicx,cite,color,soul}

\newcommand{\HB}{{\tt HiggsBounds}}
\newcommand{\HS}{{\tt HiggsSignals}}

\newcommand{\FH}{{\tt FeynHiggs}}

\newcommand{\pMSSM}{pMSSM\,8}
\newcommand{\CL}[1]{$#1~\mathrm{C.L.}$}
\newcommand{\simMH}{125}
\newcommand{\MHexp}{125}
\newcommand{\btn}{\ensuremath{\br(B_u \to \tau \nu_\tau)}}
\newcommand{\bmm}{\ensuremath{\br(B_s \to \mu^+\mu^-)}}
\newcommand{\bsg}{\ensuremath{\br(B \to X_s \ga)}}
\newcommand{\gmt}{\ensuremath{(g-2)_\mu}}

\newcommand{\MS}{M_S}

\newcommand{\lowMH}{low-$\MH$}

\newcommand{\lowMHuplow}{low-$\MH^{\rm alt-}$}
\newcommand{\lowMHuphigh}{low-$\MH^{\rm alt+}$}
\newcommand{\lowMHupvar}{low-$\MH^{\rm alt\,v}$}

\newenvironment{Eqnarray}%
         {\arraycolsep 0.14em\begin{eqnarray}}{\end{eqnarray}}
\def\beqa{\begin{Eqnarray}}
\def\eeqa{\end{Eqnarray}}
\def\beq{\begin{equation}}
\def\eeq{\end{equation}}

\def\ls#1{\ifmath{_{\lower1.5pt\hbox{$\scriptstyle #1$}}}}

\def\ifmath#1{\relax\ifmmode #1\else $#1$\fi}

\ShortTitle{The heavy Higgs case}

\author{\speaker{S. Heinemeyer}\\ %\thanks{A footnote may follow.}\\
  Campus of International Excellence UAM+CSIC, Cantoblanco, Madrid, Spain;\\
  Instituto de F\'isica Te\'orica UAM-CSIC, C/ Nicolas Cabrera 13-15,
  Madrid, Spain; \\
  Instituto de F\'isica de Cantabria (CSIC-UC), Santander, Spain\\
        E-mail: \email{Sven.Heinemeyer@cern.ch}}

\abstract{
  We review that the heavy $\cp$-even MSSM 
    Higgs boson is still a viable candidate to explain the Higgs
    signal at \simMH~GeV.
    This is possible in a highly constrained parameter region,
    that will be probed by LHC searches for the $\cp$-odd
    Higgs boson and the charged Higgs boson in the near future.
  We briefly discuss the  new benchmark scenarios that
    can be employed to maximize the sensitivity of the experimental
    analysis to this interpretation.
}

\FullConference{Prospects for Charged Higgs Discovery at Colliders\\ 
		3-6 October 2016\\
		Uppsala University, Sweden}

\include{paperdef}
\begin{document}

%%%%%%%%%%%%%%%%%%%%%%%%%%%%%%%%%%%%%%%%%%%%%%%%%%%%%%%%%%%%%%%%%%%%%%%%%%%%%%%
%%%%%%%%%%%%%%%%%%%%%%%%%%%%%%%%%%%%%%%%%%%%%%%%%%%%%%%%%%%%%%%%%%%%%%%%%%%%%%%

\section{Introduction}

The discovery of a SM-like Higgs boson in Run~I of the Large Hadron
Collider (LHC)~\cite{ATLASdiscovery,CMSdiscovery} marks a milestone in
the exploration of electroweak symmetry breaking (EWSB).  Within
experimental and theoretical uncertainties, the properties of the new
particle are compatible with the Higgs boson of the Standard Model
(SM)~\cite{ATLAS-CMS-comb}. Looking beyond the SM, also the light
$\cp$-even Higgs boson of the Minimal Supersymmetric Standard Model
(MSSM)~\cite{mssm} is a perfect candidate, as it possesses
SM Higgs-like properties over a significant part of the model parameter
space with only small deviations 
from the SM in the Higgs production and decay rates~\cite{Mh125}.

Here we will review~\cite{hifi2} that also the {\em heavy} $\cp$-even
Higgs boson of 
the MSSM is a viable candidate to explain the observed signal at \simMH~GeV.
(the ``heavy Higgs case'', which has been discussed in
\citeres{Mh125,Hagiwara:2012mga,Benbrik:2012rm,Drees:2012fb,Han:2013mga,hifi,hifi2}).
At lowest order, the Higgs sector of the MSSM can be fully specified in
terms of the 
$W$~and $Z$~boson masses, $\MW$ and $\MZ$, the $\cp$-odd Higgs boson
mass, $\MA$, and $\tb \equiv v_2/v_1$, the ratio of the two neutral Higgs
vacuum expectation values.  However, higher-order corrections are crucial for a
precise prediction of the MSSM Higgs boson properties and introduce dependences on other model parameters, see e.g.\
\citeres{mhiggsAWB,habilSH,PomssmRep} for reviews.

In the heavy Higgs case all five MSSM Higgs bosons
are relatively light, and in particular the lightest $\cp$-even
Higgs boson has a mass (substantially) smaller than $\simMH\gev$
with suppressed couplings to gauge bosons. We review whether
the heavy Higgs case in the MSSM can still provide a good theoretical
description of the current experimental data, and 
which parts of the parameter space of the MSSM are favored.
We also discuss the newly defined benchmark scenarios in which this
possibility is realized, in agreement with all current Higgs constraints.

%%%%%%%%%%%%%%%%%%%%%%%%%%%%%%%%%%%%%%%%%%%%%%%%%%%%%%%%%%%%%%%%%%%%%%%%%%%%%%%
%%%%%%%%%%%%%%%%%%%%%%%%%%%%%%%%%%%%%%%%%%%%%%%%%%%%%%%%%%%%%%%%%%%%%%%%%%%%%%%

\section{Theoretical basis}
\label{sec:theory}

In the supersymmetric extension of the SM, an even number of Higgs
multiplets consisting of pairs of Higgs doublets with opposite
hypercharge is required to avoid anomalies due to the supersymmetric
Higgsino partners. Consequently the MSSM employs two Higgs doublets,
denoted by $H_1$ and $H_2$, with hypercharges $-1$ and $+1$,
respectively.   After minimizing the scalar potential, the neutral
components of $H_1$ and $H_2$ acquire vacuum expectation values (vevs),
$v_1$ and $v_2$. Without loss of generality, one can assume that the
vevs are real and non-negative, yielding
\begin{align}
v^2\equiv v_1^2+v_2^2\simeq(246~{\rm GeV})^2\,, \quad
\tb\equiv v_2/v_1\,.
\end{align}
The two-doublet Higgs sector gives rise to five physical Higgs states.
Neglecting $\cp$-violating phases the mass eigenstates correspond to the
neutral $\cp$-even Higgs bosons $h$, $H$ (with $M_h<M_H$), the $\cp$-odd
$A$, and the charged Higgs pair $H^\pm$.  

At lowest order,  the MSSM Higgs sector is fully described by 
$\MZ$ and two MSSM parameters, conveniently chosen as $\MA$, and $\tb$.
Higher order corrections to the Higgs masses are known to be sizable
and must be included, in order to be consistent with the observed Higgs
signal at $\MHexp \gev$~\cite{ATLAS-CMS-comb}.
In order to shift the mass of $h$ up to $\MHexp \gev$,
large radiative corrections are necessary, which
require a large splitting in the stop sector and/or heavy stops.
The stop (sbottom) sector is governed by the soft SUSY-breaking mass
parameter $\MstL$ and $\MstR$ ($\MsbL$ and $\MsbR$), where SU(2) gauge
invariance requires $\MstL=\MsbL$, 
the trilinear coupling $A_t$ ($A_b$) and the Higgsino mass parameter $\mu$.

%%%%%%%%%%%%%%%%%%%%%%%%%%%%%%%%%%%%%%%%%%%%%%%%%%%%%%%%%%%%%%%%%%%%%%%%%%%%%%%

%\subsection{The heavy Higgs case}
%\label{Sec:heavyHiggs}

The ``heavy Higgs case'', i.e.\ the heavy $\cp$-even Higgs boson gives
rise to the signal observed at $\simMH \gev$ can {\em only} be realized
in the \emph{alignment without decoupling limit}. 
In the so-called \textit{Higgs basis} (see \citere{hifi2} for details
and citations), the scalar Higgs potential in terms of the Higgs basis
fields $\mathcal{H}_1$ and~$\mathcal{H}_2$, can be expressed as
\begin{align}
{\cal V} = \ldots + \edz Z_1 (\cHe^\dagger\cHe)^2 + \ldots +
\KKL 
 Z_5 (\cHe^\dagger\cHz)^2 +
Z_6 (\cHe^\dagger\cHe)(\cHe^\dagger\cHz) + {\rm h.c.} \KKR
+ \ldots\,,
\end{align}
where the most important terms of the scalar potential are highlighted
above.  The quartic couplings $Z_1$, $Z_5$ and $Z_6$ are linear
combinations of the quartic couplings that appear in the MSSM Higgs
potential expressed in terms of $H_1$ and $H_2$.  
The $Z_i$ are $\mathcal{O}(1)$ parameters.

The mass matrix of the neutral $\cp$-even Higgs bosons is then given by
\begin{align}
  \cM^2 = \ML Z_1 v^2 & Z_6 v^2 \\
        Z_6 v^2 & \MA^2 + Z_5 v^2 \MR\,.
\label{HiBa-massmatrix}
\end{align}
The \emph{alignment without decoupling limit} is reached for
$|Z_6|\ll 1$.  In this case $h$ is SM-like if $\MA^2+(Z_5-Z_1)v^2>0$ and
$H$ is SM-like if $\MA^2+(Z_5-Z_1)v^2<0$: the ``heavy Higgs case''.

The possibility of alignment without decoupling has been analyzed in detail in
Refs.~\cite{Gunion:2002zf,Craig:2013hca,Carena:2013ooa,Haber:2013mia,Carena:2014nza,Dev:2014yca,Bernon:2015qea,Bernon:2015wef} (see also the
``$\tau$-phobic'' benchmark scenario in \citere{Carena:2013ytb}). It was
pointed out that exact alignment via $|Z_6| \ll 1$ can only happen through an
accidental cancellation of the tree-level terms with contributions arising
at the one-loop level (or higher).

%%%%%%%%%%%%%%%%%%%%%%%%%%%%%%%%%%%%%%%%%%%%%%%%%%%%%%%%%%%%%%%%%%%%%%%%%%%%%%%
%%%%%%%%%%%%%%%%%%%%%%%%%%%%%%%%%%%%%%%%%%%%%%%%%%%%%%%%%%%%%%%%%%%%%%%%%%%%%%%

\section{Parameter scan and observables}

The results shown below have been obtained by scanning the MSSM
parameter space. To achieve a good sampling of the full MSSM parameter
space with \order{10^7} points, we restrict ourselves to the eight 
MSSM parameters, called the \pMSSM, 
\begin{align}
\tb, \quad M_A, \quad \msqd, \quad \Af, \quad \mu, \quad \msld, \quad \mslez, \quad M_2\,,
\label{Eq:fitparameters}
\end{align}
most relevant for the phenomenology of the Higgs sector. Here $\mu$ denotes
the Higgs mixing parameter, $\msld$ ($\mslez$) is the diagonal soft
SUSY-breaking parameters for scalar leptons in the thrid (second and
first) generation, and $M_2$ denotes the SU(2) gaugino soft
SUSY-breaking parameter.
The scan assumes furthermore that the third generation squark and
slepton parameters are universal.  That is, we take 
$\msqd := \MstL  (= \MsbL) = \MstR = \MsbR$,  
$\msld := \MstauL = \MstauR = M_{\tilde \nu_\tau}$
and $\Af := \At = \Ab = \Atau$.
The remaining MSSM parameters are fixed,
\begin{align}
\MsqL = \MsqR~(q = c, s, u, d) \; &= \; 1500 \gev, \\
M_3 = \mgl &= 1500 \gev\,. 
\end{align}
The high values for the squark and gluino mass parameters, which have a
minor impact on the Higgs sector, are chosen in order to be in agreement
with the limits from direct SUSY searches. The U(1) gaugino mass
parameter is fixed via the usual GUT relation.
The \pMSSM\ parameter space is scanned with uniformly distributed random
values in the eight input parameters
over the parameter ranges given in \refta{tab:param}.

%%%T A B L E %%%%%%%%%%%%%%%%%%%%%%%%%%%%%%%%%%%%%%%%%%%%%%%%%%%%%
\begin{table}[h!]
\centering
\begin{tabular}{|r|cc|}
\hline
Parameter &  Minimum &  Maximum \\
\hline
$\MA$ [GeV]    & 90 & 200 \\
$\tb$ \phantom{[GeV]}        &1 & 20 \\
$M_{\tilde{q}_3}$ [GeV]  & 200      & 1500 \\
$\msld$ [GeV]  & 200      & 1000 \\
$\mslez$ [GeV] & 200      & 1000 \\
$\mu$ [GeV]    & $-5000$ & 5000 \\
$\Af$ [GeV]    & $-3\,M_{\tilde{q}_3}$ & $3\,M_{\tilde{q}_3}$\\
$M_2$ [GeV]    & 200      & 500 \\
\hline
\end{tabular}
\caption{Ranges used for the free parameters in the \pMSSM\ scan.}
\label{tab:param}
\end{table}
%%%T A B L E %%%%%%%%%%%%%%%%%%%%%%%%%%%%%%%%%%%%%%%%%%%%%%%%%%%%%

We calculate the SUSY particle spectrum and the MSSM Higgs masses using 
\FH\ (version 2.11.2)%
\footnote{
Recent updates in the Higgs boson mass calculations~\cite{FHwww} lead to a
downward shift in $\Mh$, in particular for large values of $X_t/M_S$. These
changes range within the estimated uncertainties and should not have a
drastic impact on our analysis.}%
~\cite{FHwww,feynhiggs,mhiggsAEC,mhcmssmlong}, 
and estimate the remaining theoretical uncertainty (e.g.~from unknown higher-order corrections) in the Higgs mass calculation 
to be $3\gev$~\cite{mhiggsAEC}.
Following \citeres{Benbrik:2012rm,hifi}, we demand that all points
fulfill a $\matr{Z}$-matrix criterion, 
$\left||Z_{21}^{\mathrm{2L}}| - |Z_{21}^{\mathrm{1L}}|\right|/
|Z_{21}^{\mathrm{1L}}|<0.25$
in order to ensure a reliable and stable perturbative behavior in the 
calculation of propagator-type contributions in the MSSM Higgs sector.
The $\matr{Z}$-matrix definition and details can be found in
\citere{mhcmssmlong}. 

The observables included in the fit are the Higgs-boson mass, the Higgs
signal rates (evaluated with
\HS~\cite{higgssignals}),
Higgs exclusion bounds from LEP, Tevatron and the LHC (evaluated with
\HB~\cite{higgsbounds}),
SUSY exclusion bounds from the LEP and the LHC (the latter evaluated with
{\tt CheckMate}~\cite{checkmate}), 
and several low-energy observables (LEOs): \bsg, \bmm\ and \btn\ (evaluated with
{\tt SuperIso}~\cite{superiso}),
\gmt (evaluated with {\tt SuperIso} and \FH), and
$\MW$ (with an evaluation based on \citere{mw}). The total $\chi^2$ is
evaluated as (see \citere{hifi2} for more details), 
\begin{align}
\chi_H^2 &= \frac{(M_{H}-\MHhat)^2}{\sigma_{\MHhat}^2}
         + \chi^2_\text{HS}
         +\sum_{i=1}^{n_{\mathrm{LEO}}} \frac{(O_i-\hat{O}_i)^2}{\sigma_i^2} 
         - 2\ln\mathcal{L}_\text{limits}~,
\label{eq:totchi2}
\end{align}
where experimental measurements are denoted with a hat.

%%%%%%%%%%%%%%%%%%%%%%%%%%%%%%%%%%%%%%%%%%%%%%%%%%%%%%%%%%%%%%%%%%%%%%%%%%%%%%%
%%%%%%%%%%%%%%%%%%%%%%%%%%%%%%%%%%%%%%%%%%%%%%%%%%%%%%%%%%%%%%%%%%%%%%%%%%%%%%%

\section{Results for the ``heavy Higgs case''}

Based on the above described $\chi^2$ evaluation the best-fit point,
shown as a star below, and the preferred parameter regions are
derived. Points with $\Delta\chi_H^2 < 2.30~(5.99)$ are highlighted in
red (yellow), corresponding to points in a two-dimensional ${68}\%$
($95\%$) C.L.~region in the Gaussian limit. The best fit point has
a $\chi^2/$dof of $73.7/85$, corresponding to a $p$-value of $0.87$,
i.e.\ the heavy Higgs case presents an excellent fit to the experimental
data~\cite{hifi2}.

\begin{figure}[htb!]
\includegraphics[width=0.46\columnwidth]{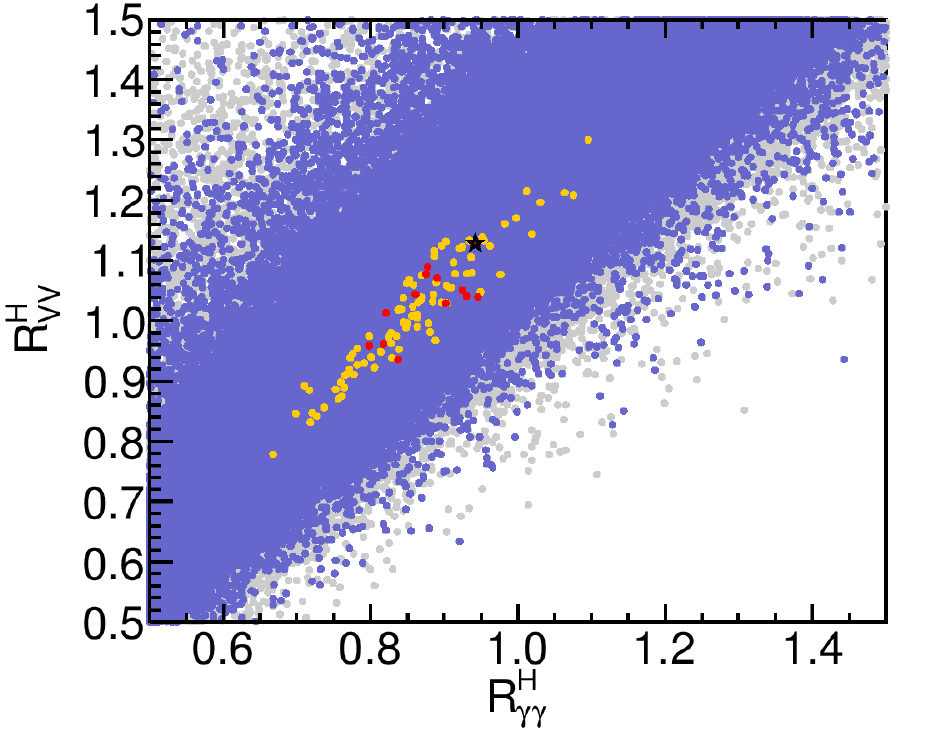}\hspace{0.5cm}
\includegraphics[width=0.46\columnwidth]{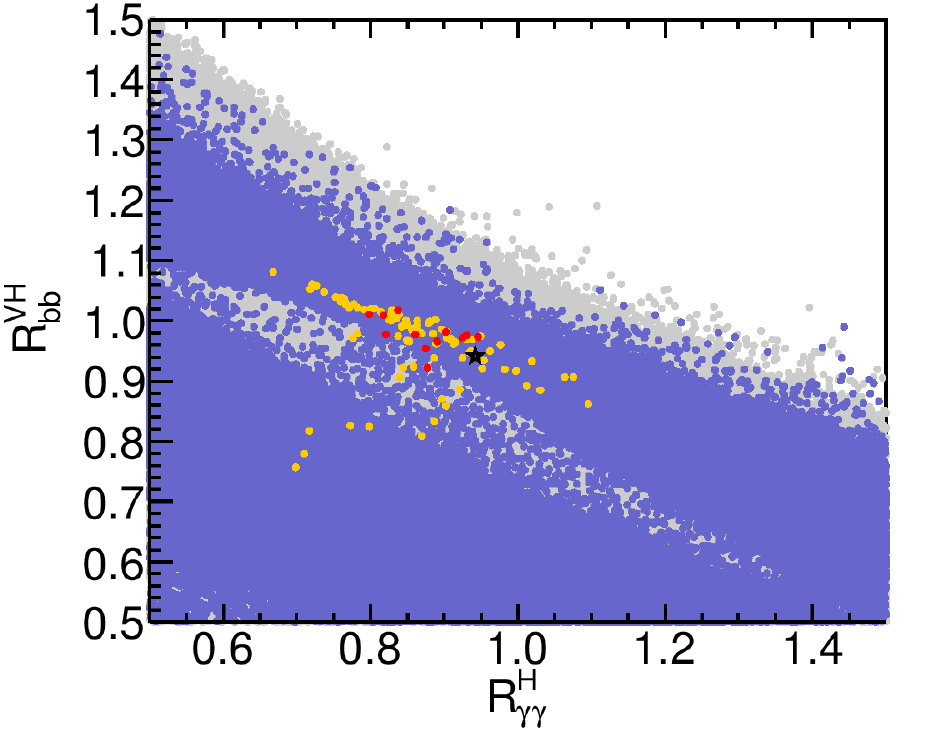}\\
\includegraphics[width=0.46\columnwidth]{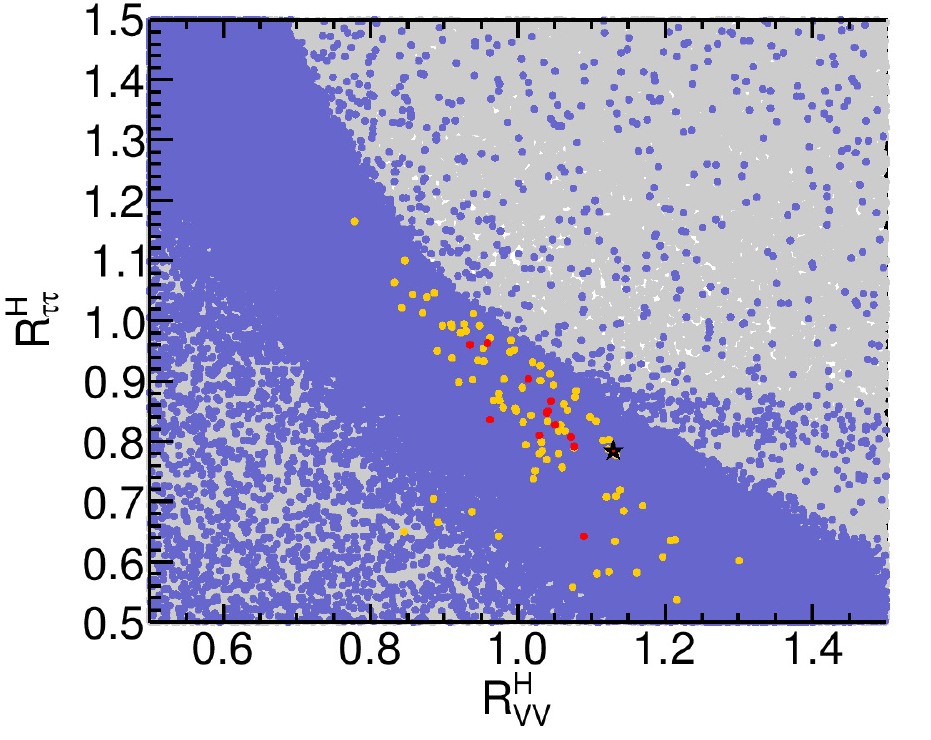}\hspace{0.5cm}
\includegraphics[width=0.46\columnwidth]{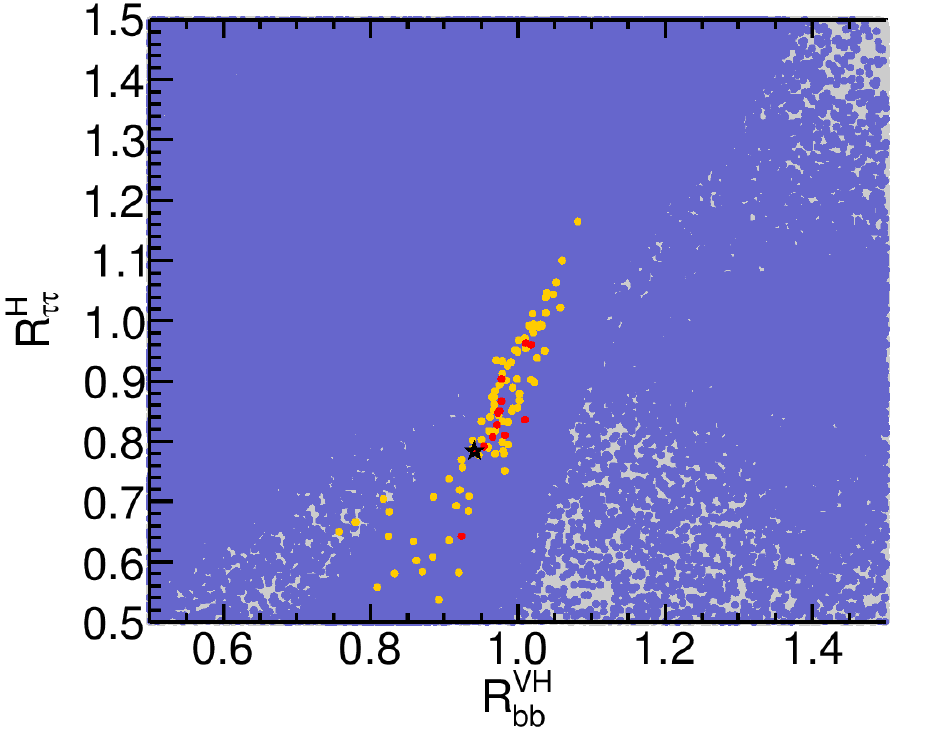}
\caption{Correlations between signal rates for the
  heavy Higgs case. The best-fit points are shown as a black star, 
  and points with $\Delta \chi_H^2 < 2.3$ (shown in \emph{red})
  and $\Delta \chi_H^2 < 5.99$ (shown in \emph{yellow}). }
\label{fig:Hrates_corr}
\end{figure}

In \reffi{fig:Hrates_corr}~\cite{hifi2} we review the correlations for
the heavy Higgs signal rates,
\begin{align}
R_{XX}^{P(H)} = \frac{\sum_{P(H)} \sigma(P(H)) \times \br(H\to XX)}{\sum_{P(H)} \sigma_\mathrm{SM}(P(H)) \times \br_\mathrm{SM}(H\to XX)}.
\label{Eq:Rvalues}
\end{align}
Here $XX= VV, \gamma\gamma, bb, \tau\tau$ (with $V=W^\pm,Z$) denotes the
final state from the Higgs decay and $P(H)$
denotes the Higgs production mode. It can be seen that the heavy Higgs
case can reproduce the SM case ($R_{XX}^{P(H)} = 1$), but also allows
for some spread, in particular in $R_{\tau\tau}^{H}$. 

%%% F I G U R E %%%%%%%%%%%%%%%%%%%%%%%%%%%%%%%%%%%%%%%%%%%%%%%%%%%%%
%%% F I G U R E %%%%%%%%%%%%%%%%%%%%%%%%%%%%%%%%%%%%%%%%%%%%%%%%%%%%%

%%% F I G U R E %%%%%%%%%%%%%%%%%%%%%%%%%%%%%%%%%%%%%%%%%%%%%%%%%%%%%
\begin{figure}[htb!]
%\vspace{0.5em}
\includegraphics[width=0.46\columnwidth]{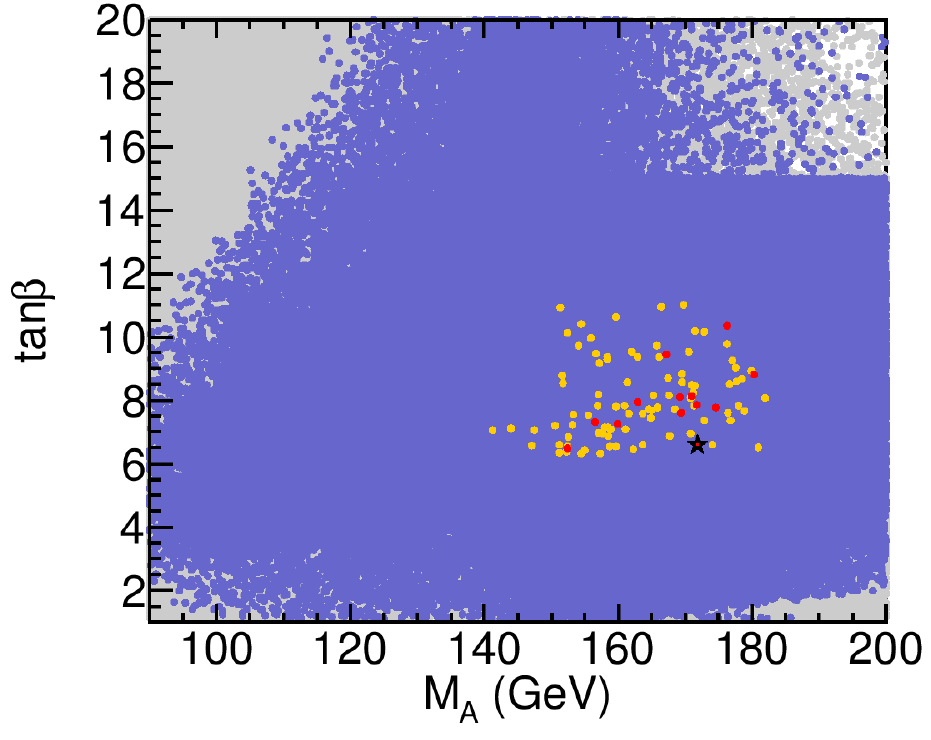}\hspace{0.5cm}
\includegraphics[width=0.46\columnwidth]{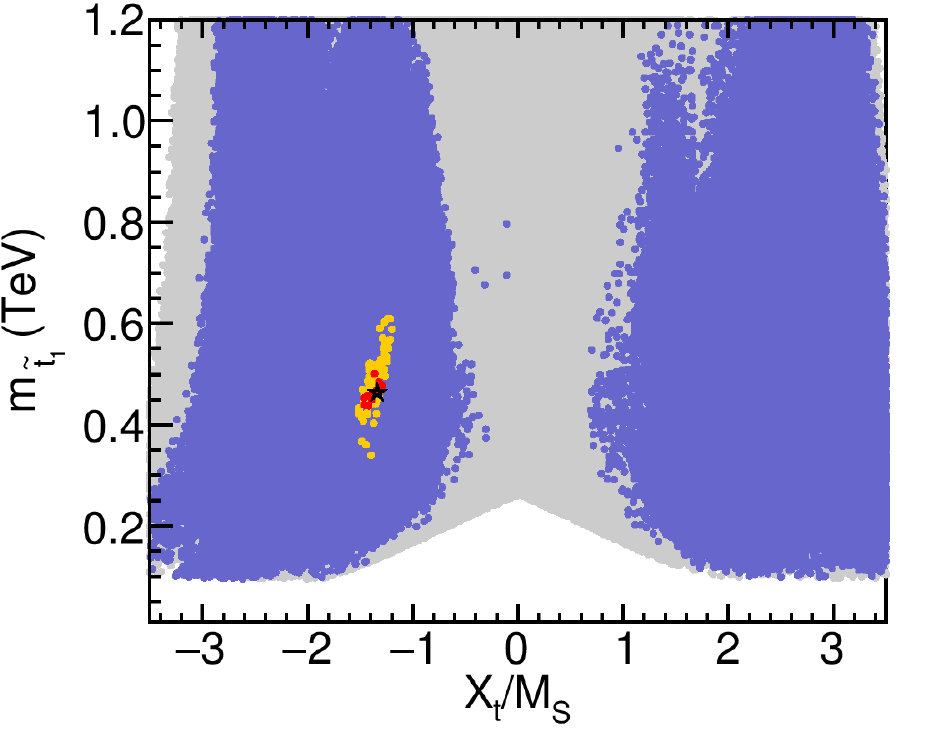}
\caption{$\MA$-$\tb$ plane (left) and $\Xt/\MS$-$\mste$ plane (right)
  in the heavy Higgs case. The color coding is as in Fig.~1.}
\label{fig:MAtb}
\end{figure}
%%% F I G U R E %%%%%%%%%%%%%%%%%%%%%%%%%%%%%%%%%%%%%%%%%%%%%%%%%%%%%

\begin{figure}[htb!]
\includegraphics[width=0.46\columnwidth]{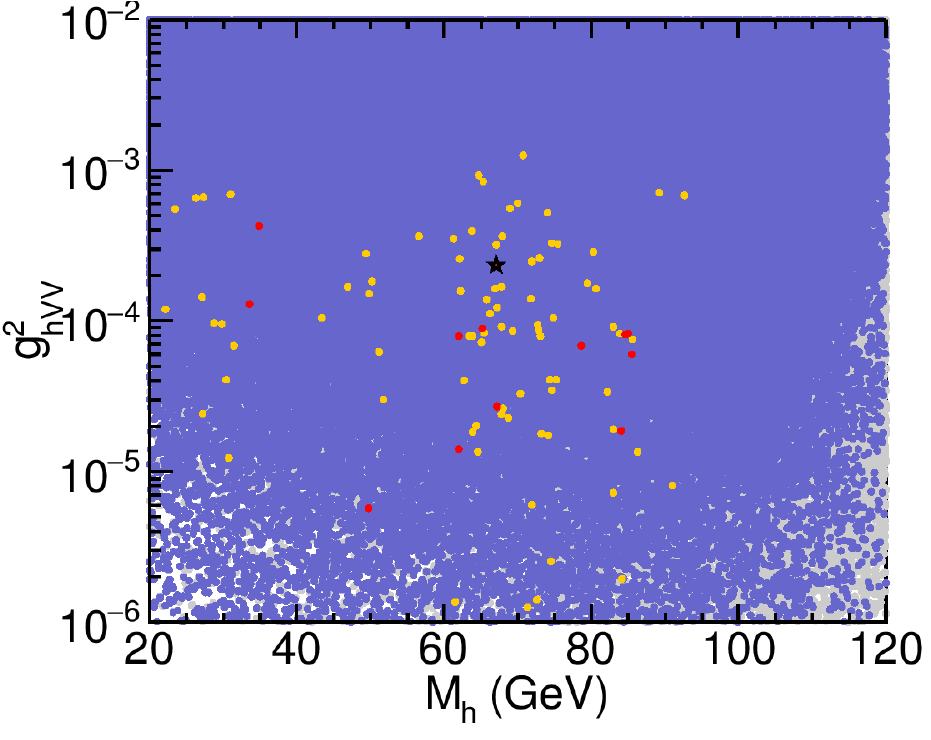}\hspace{0.5cm}
\includegraphics[width=0.46\columnwidth]{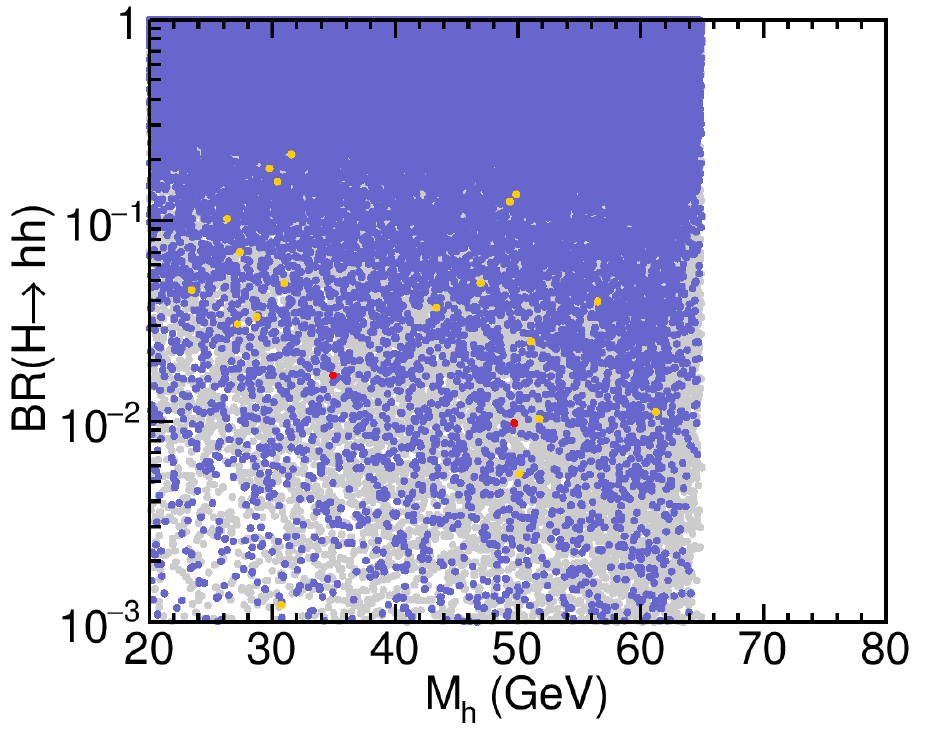}
\caption{$g_{hVV}^2$ (relative to the SM value) 
  (left) and $\br(H \to hh)$ as a function of $\Mh$ (right)
  in the heavy Higgs case. The color coding is as in Fig~1.}
\label{fig:h}
\end{figure}

The MSSM parameter space for the heavy Higgs scenario is shown in
\reffi{fig:MAtb}. The left plot indicates the preferred regions in the
$\MA$-$\tb$ plane, where one can see that
$140 \gev \lsim \MA \lsim 185 \gev$ must be fulfilled, while $\tb$
ranges between $\sim 6$ and $\sim 11$. The right plot shows the
preferred regions in the $\Xt/\MS$-$\mste$ plane. Here the heavy Higgs
case makes a clear prediction with $300 \gev \lsim$ $\mste \lsim 650 \gev$
and $\Xt/\MS \sim -1.5$. Some properties of the light $\cp$-even
Higgs boson are shown in \reffi{fig:h}. The left plot shows the light
Higgs boson coupling to massive gauge bosons relative to the SM
value. One can see that the coupling squared is suppressed by a factor
of 1000 or more, rendering its discovery via $e^+e^- \to Z^* \to Zh$
at LEP impossible~\cite{LEPHiggsSM,LEPHiggsMSSM}. The right plot gives
the $\br(H \to hh)$ for $\Mh \lsim \MH/2$. Here it is shown that the
BR does not exceed 20\%, and thus does not distort the coupling
measurements of the heavy Higgs at $\sim 125 \gev$ too
much~\cite{ATLAS-CMS-comb}. 

%%% F I G U R E %%%%%%%%%%%%%%%%%%%%%%%%%%%%%%%%%%%%%%%%%%%%%%%%%%%%%
%%% F I G U R E %%%%%%%%%%%%%%%%%%%%%%%%%%%%%%%%%%%%%%%%%%%%%%%%%%%%%

%%%%%%%%%%%%%%%%%%%%%%%%%%%%%%%%%%%%%%%%%%%%%%%%%%%%%%%%%%%%%%%%%%%%%%%%%%%%%%%
%%%%%%%%%%%%%%%%%%%%%%%%%%%%%%%%%%%%%%%%%%%%%%%%%%%%%%%%%%%%%%%%%%%%%%%%%%%%%%%

\section{Updated benchmark scenarios}

In \citere{hifi2} an updated set of benchmarks for the heavy Higgs
case was presented, superseeding the experimentally excluded
\lowMH\ scenario~\cite{Carena:2013ytb}. The parameters of the three
new benchmark scenarios are given in \refta{tab:benchmarks}. 
The \lowMHuplow\ (\lowMHuphigh) scenario is defined in the $\mu$-$\tb$
plane with $\MHp < (>) \mt$, while the \lowMHupvar\ scenario has a
fixed $\mu$ in the $\MHp$-$\tb$ plane.

%%%T A B L E %%%%%%%%%%%%%%%%%%%%%%%%%%%%%%%%%%%%%%%%%%%%%%%%%%%%%
\begin{table}[h!]
\centering
\begin{tabular}{lccc}
\hline
Benchmark scenario		& 	$\MHp~[\mathrm{GeV}]$ & $\mu~[\mathrm{GeV}]$ & $\tb$ \\
\hline
\lowMHuplow &	$155$		&	$3800$ -- $6500$	&		$4$ -- $9$ \\
\lowMHuphigh &	$185$		&	$4800$ -- $7000$	&		$4$ -- $9$ \\
\lowMHupvar  &		$140$ -- $220$ &     $6000$			&		$4$ -- $9$ \\
\hline
fixed parameters: &\multicolumn{3}{l}{$\mt = 173.2\gev$, \quad $\At = \Atau = \Ab = -70\gev$,\quad $M_2 = 300 \gev$,}\\ 
&\multicolumn{3}{l}{$\MsqL = \MsqR =  1500 \gev$~($q = c, s, u, d$),\quad $\mgl = 1500 \gev$,} \\
&\multicolumn{3}{l}{$M_{\tilde{q}_3} = 750\gev$,\quad$\mslez = 250 \gev$,\quad $\msld = 500 \gev$} \\
\hline
\end{tabular}
\caption{Parameters of the updated \lowMH\ benchmark scenarios, see
  \citere{hifi2} for more details.
The lower row gives the fixed parameters that are common to all three
benchmark scenarios
and $M_1 = \tfrac{5}{3} \tfrac{\sw^2}{\cw^2} M_2$}.
\label{tab:benchmarks}
\end{table}
%%%T A B L E %%%%%%%%%%%%%%%%%%%%%%%%%%%%%%%%%%%%%%%%%%%%%%%%%%%%%

\begin{figure}[ht!]
\centering
\includegraphics[width=0.45\columnwidth]{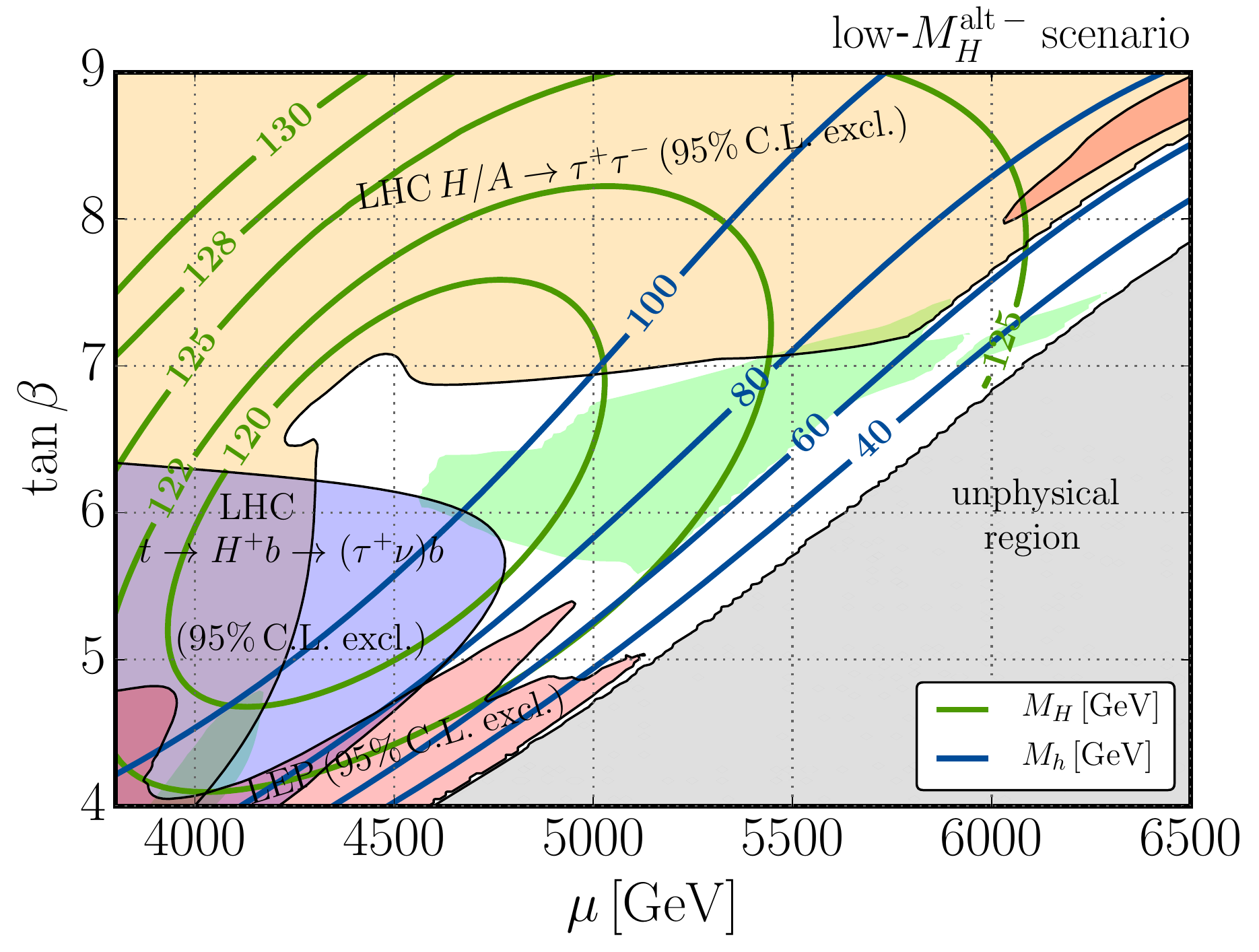}\hfill
\includegraphics[width=0.45\columnwidth]{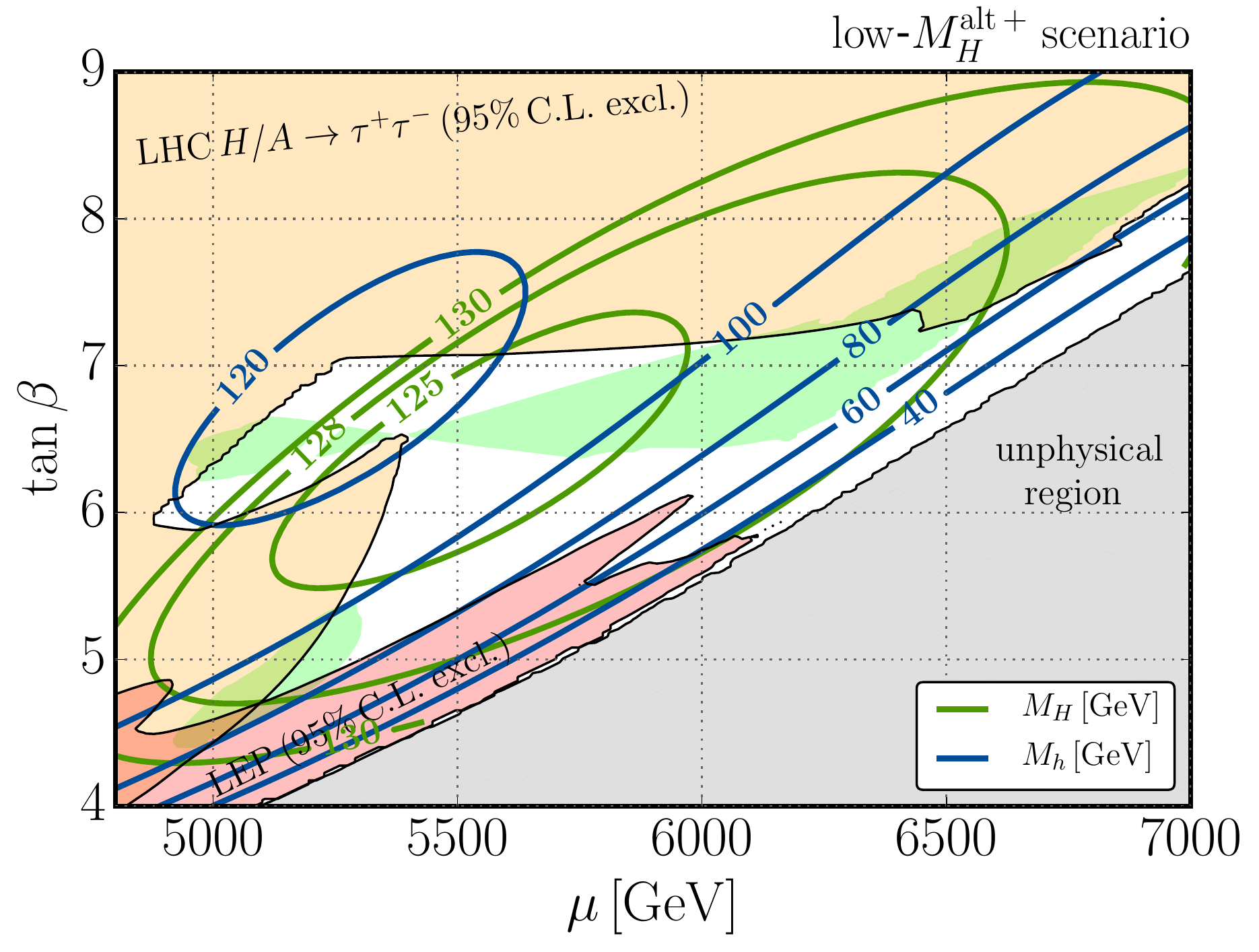}
\includegraphics[width=0.45\columnwidth]{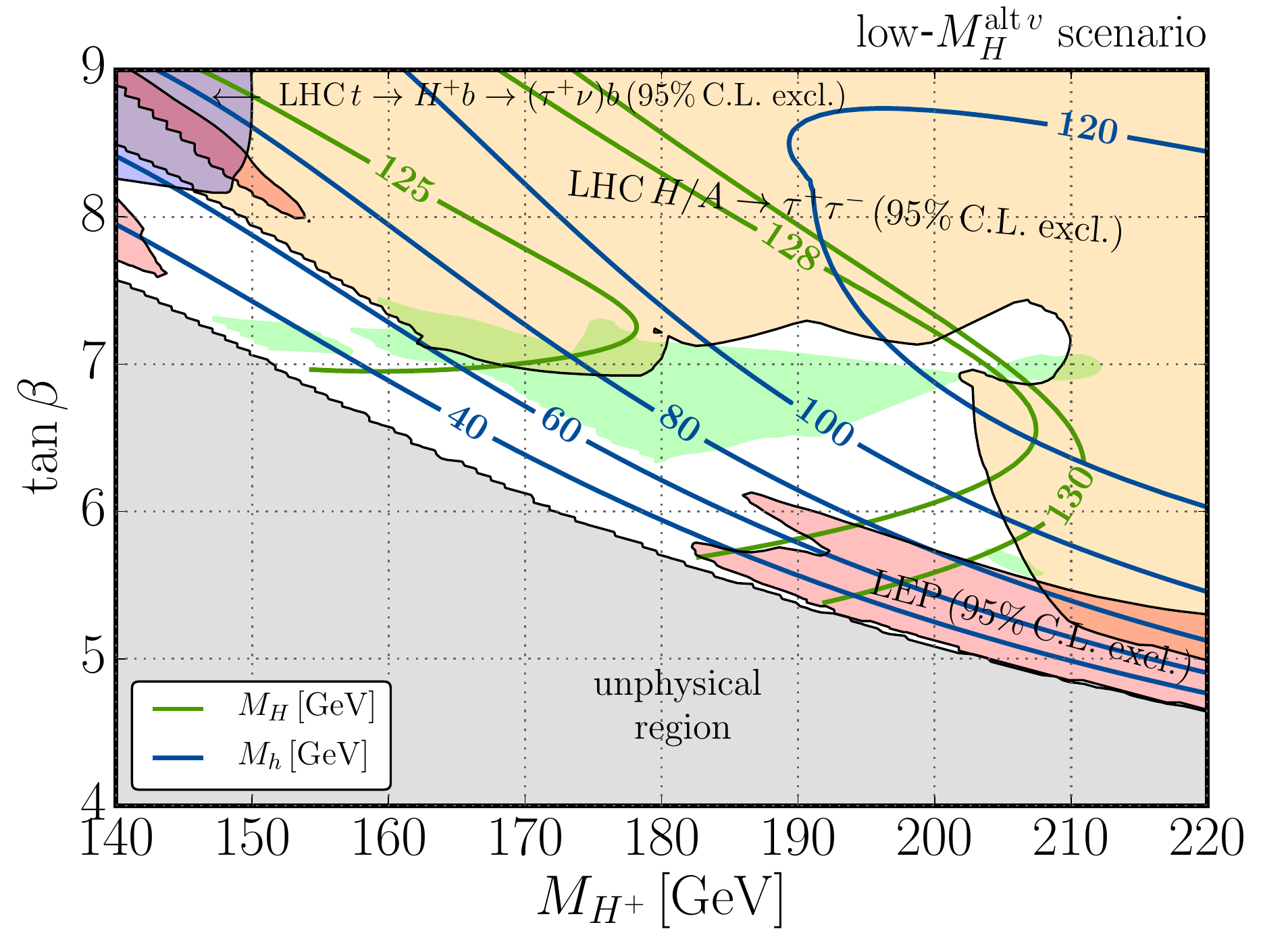}
\caption{The \lowMHuplow\ and \lowMHuphigh\ benchmark scenarios in
  the $\mu$-$\tb$ plane with $\MHp = 155\gev$ (\emph{upper left}),
  and with $\MHp = 185\gev$ (\emph{lower right}),
  and the \lowMHupvar\  benchmark scenario in the $\MHp$-$\tb$ plane
  with $\mu = 6000 \gev$ in the lower row. For the color coding and
  line styles see text.
}  
\label{fig:lowMH}
\end{figure}

The experimentally allowed parameter space in the three benchmark
scenarios is shown in \reffi{fig:lowMH}.%
\footnote{In the evaluation of these plots the two-loop corrections to
  the $\MA$-$\MHp$ mass relation had been omitted. Taking them into
  account will lead to a slight shift of the $\Mh$ contour lines.}%
~The red,
orange and blue regions are disfavoured at the \CL{95\%} by LEP light
Higgs $h$ searches~\cite{LEPHiggsMSSM}, LHC $H/A\to \tau^+\tau^-$
searches~\cite{Khachatryan:2014wca,CMS:2015mca} and LHC $t\to
H^+b\to(\tau\nu)b$ searches~\cite{Aad:2014kga,Khachatryan:2015qxa},
respectively. The green area indicates parameter regions that are
compatible with the Higgs signal (at $\sim$ \CL{95\%}, see \citere{hifi2} for
details), unphysical regions are displayed in gray.
Contour lines indicate the Higgs masses $M_h$ and $M_H$ (in
GeV).

%% F I G U R E %%%%%%%%%%%%%%%%%%%%%%%%%%%%%%%%%%%%%%%%%%%%%%%%%%%%%
%% F I G U R E %%%%%%%%%%%%%%%%%%%%%%%%%%%%%%%%%%%%%%%%%%%%%%%%%%%%%

While being ``squeezed'' from different searches, \reffi{fig:lowMH}
shows that the heavy Higgs case remains a valid option with the
interesting feature of a light $\cp$-even Higgs {\em below} $125 \gev$. We
hope that the new benchmark scenarios facilitate the search for these
light Higgs bosons as well as for the heavier, not yet discovered Higgs bosons
in Run~II.

%%%%%%%%%%%%%%%%%%%%%%%%%%%%%%%%%%%%%%%%%%%%%%%%%%%%%%%%%%%%%%%%%%%%%%%%%%%%%%%
%%%%%%%%%%%%%%%%%%%%%%%%%%%%%%%%%%%%%%%%%%%%%%%%%%%%%%%%%%%%%%%%%%%%%%%%%%%%%%%

\newpage
\section{Conclusinos}

We have briefly reviewed the case that the Higgs boson observed at
$\sim 125 \gev$ is the heavy $\cp$-even Higgs boson in the MSSM, as
recently analyzed in \citere{hifi2}. The analysis uses an
eight-dimensional MSSM parameter scan to find the regions in the
parameter space that fit best the experimental data. It was found that
the rates of the heavy $\cp$-Higgs boson are close to the SM rates,
but can still differ by 20\% or more to yield a good fit. Parameters
such as $\MA$, $\tb$ or $\mste$ are confined to relatively small
intervals, making clear predictions for Higgs and SUSY searches. The light
$\cp$-even Higgs boson escaped the LEP searches via a tiny coupling to
SM gauge bosons, and the decay $H \to hh$ is sufficiently suppressed
not to impact too strongly the heavy Higgs boson rates. Three new
benchmark scenarios have been reviewed that have been defined to
facilitate the experimental searches at the LHC Run~II.

%%%%%%%%%%%%%%%%%%%%%%%%%%%%%%%%%%%%%%%%%%%%%%%%%%%%%%%%%%%%%%%%%%%%%%%%%%%%%%%

\subsection*{Acknowledgements}

I thank
P.~Bechtle,
H.~Haber,
O.~St{\aa}l,
T.~Stefaniak, 
G.~Weiglein and
L.~Zeune, with whom the results shown here have been derived.
I~furthermore thank S.~Pa\ss ehr for helpful discussions.
I~thank the organizers of C$H^{\mbox{}^\pm}$\hspace{-2.5mm}arged 2016
for the invitation and the, as always, pleasant and productive
atmosphere, as well as for financial support.
The work of S.H.\ is supported in part by CICYT 
(grant FPA 2013-40715-P) and by the Spanish MICINN's Consolider-Ingenio 
2010 Program under grant MultiDark CSD2009-00064.

%%%%%%%%%%%%%%%%%%%%%%%%%%%%%%%%%%%%%%%%%%%%%%%%%%%%%%%%%%%%%%%%%%%%%%%%%%%%%%%
%%%%%%%%%%%%%%%%%%%%%%%%%%%%%%%%%%%%%%%%%%%%%%%%%%%%%%%%%%%%%%%%%%%%%%%%%%%%%%%

\end{document}

%%%%%%%%%%%%%%%%%%%%%%%%%%%%%%%%%%%%%%%%%%%%%%%%%%%%%%%%%%%%%%%%%%%%%%%%%%%%%%

%1
\bibitem{higgs-mechanism}
        P.~Higgs,
        {\em Phys.\ Lett.} {\bf 12} (1964) 132;
        %%CITATION = PHLTA,12,132;%%
        {\em Phys.\ Rev.\ Lett.} {\bf 13} (1964) 508;
        %%CITATION = PRLTA,13,508;%%
        {\em Phys.\ Rev.} {\bf 145} (1966) 1156;
        %%CITATION = PHRVA,145,1156;%%
        F.~Englert and R.~Brout,
        {\em Phys.\ Rev.\ Lett.} {\bf 13} (1964) 321;
        %%CITATION = PRLTA,13,321;%%
        G.~~Guralnik, C.~~Hagen and T.~Kibble,
        {\em Phys.\ Rev.\ Lett.} {\bf 13} (1964) 585.
        %%CITATION = PRLTA,13,585;%%

%4
\bibitem{TevHiggsfinal} CDF Collaboration, D\O\ Collaboration, 
  [arXiv:1207.0449 [hep-ex]].
  %%CITATION = ARXIV:1207.0449;%

%5
\bibitem{ATLAS-Higgs-WWW} ATLAS Collaboration, see:\\
  {\tt https://twiki.cern.ch/twiki/bin/view/AtlasPublic/HiggsPublicResults}~.

%6
\bibitem{CMS-Higgs-WWW} CMS Collaboration, see:\\
  {\tt https://twiki.cern.ch/twiki/bin/view/CMSPublic/PhysicsResultsHIG}~.

%7
\bibitem{sm} S.~Glashow, 
             {\em Nucl.\ Phys.} {\bf B 22} (1961) 579; 
             %%CITATION = NUPHA,22,579;%%
             S. Weinberg, 
             {\em Phys. Rev. Lett.} {\bf 19} (1967) 19; 
             %%CITATION = PRLTA,19,1264;%%
             A. Salam, in: {\em Proceedings of the 8th Nobel 
             Symposium}, Editor N. Svartholm, Stockholm, 1968.

%8
\bibitem{mssm} H.~Nilles, 
               {\em Phys.\ Rept.} {\bf 110} (1984) 1; 
               %%CITATION = PRPLC,110,1;%%
               H.~Haber and G.~Kane, 
               {\em Phys.\ Rept.} {\bf 117} (1985) 75; 
               %%CITATION = PRPLC,117,75;%%
               R.~Barbieri, 
               {\em Riv.\ Nuovo Cim.} {\bf 11} (1988) 1. 
               %%CITATION = RNCIB,11,1;%%

%9
\bibitem{Mh125} S.~Heinemeyer, O.~St{\aa}l and G.~Weiglein, 
                {\em Phys.\ Lett.} {\bf B 710} (2012) 201
                [arXiv:1112.3026 [hep-ph]]; 
                %%CITATION = ARXIV:1112.3026;%%

%10
\bibitem{thdm} S.~Weinberg,
        Phys.\ Rev.\ Lett.\  {\bf 37} (1976) 657;
        %%CITATION = PRLTA,37,657;%%
        J.~Gunion, H.~Haber, G.~Kane and S.~Dawson,
        {\em The Higgs Hunter's Guide}
        (Perseus Publishing, Cambridge, MA, 1990),
        and references therein;
        G.~Branco et al., 
        {\em Phys.\ Rept.} {\bf 516} (2012) 1
        [arXiv:1106.0034 [hep-ph]].
        %%CITATION = ARXIV:1106.0034;%%

%11
\bibitem{thdm-types} V.~Barger, J.~Hewett and R.~Phillips,
  {\em Phys.\ Rev.} {\bf D 41} (1990) 3421.
  %%CITATION = PHRVA,D41,3421;%%

%12
\bibitem{NMSSM-etc}
        P.~Fayet, 
        {\em Nucl. Phys.} {\bf B 90} (1975) 104;
        {\em Phys. Lett.} {\bf B 64} (1976) 159;
        {\em Phys. Lett.} {\bf B 69} (1977) 489;
        {\em Phys. Lett.} {\bf B 84} (1979) 416;
        H.P.~Nilles, M.~Srednicki and D.~Wyler,
        {\em Phys. Lett.} {\bf B 120} (1983) 346;
        J.M.~Frere, D.R.~Jones and S.~Raby,
        {\em Nucl. Phys.} {\bf B 222} (1983) 11;
        J.P.~Derendinger and C.A.~Savoy, 
        {\em Nucl. Phys.} {\bf B 237} (1984) 307;
        J.~Ellis, J.~Gunion, H.~Haber, L.~Roszkowski and F.~Zwirner,
        {\em Phys. Rev.} {\bf D 39}  (1989) 844;
        M.~Drees, 
        {\em Int. J. Mod. Phys.} {\bf A 4}  (1989) 3635;
        U.~Ellwanger, C.~Hugonie and A.~Teixeira,
        {\em Phys.\ Rept.} {\bf 496} (2010) 1 
        [arXiv:0910.1785 [hep-ph]];
        %%CITATION = ARXIV:0910.1785;%%
        M.~Maniatis,
        {\em Int.\ J.\ Mod.\ Phys.} {\bf A 25} (2010) 3505
        [arXiv:0906.0777 [hep-ph]].
        %%CITATION = ARXIV:0906.0777;%%

%13
\bibitem{triplet} H.~Georgi and M.~Machacek,
  {\em Nucl.\ Phys.} {\bf B 262} (1985) 463;
  %%CITATION = NUPHA,B262,463;%%
  M.~Chanowitz and M.~Golden,
  {\em Phys.\ Lett.} {\bf 165} (1985) 105.
  %%CITATION = PHLTA,B165,105;%%

%14
\bibitem{LHCHXSWG-www1} 
 {\tt https://twiki.cern.ch/twiki/bin/view/LHCPhysics/CrossSections2011}~.

%15
\bibitem{YR1} S.~Dittmaier et al. 
              [LHC Higgs Cross Section Working Group],
              arXiv:1101.0593 [hep-ph].
              %%CITATION = ARXIV:1101.0593;%%

%16
\bibitem{YR2} S.~Dittmaier et al. 
              [LHC Higgs Cross Section Working Group],
              arXiv:1201.3084 [hep-ph].
              %%CITATION = ARXIV:1201.3084;%%

%17
\bibitem{YR3} S.~Heinemeyer et al.
              [LHC Higgs Cross Section Working Group],
              arXiv:1307.1347 [hep-ph].
              %%CITATION = ARXIV:1307.1347;%%

%18
\bibitem{HiggsRecommendation} LHC Higgs Cross Section Working Group,
  A.~David et al., 
  arXiv:1209.0040 [hep-ph].
  %%CITATION = ARXIV:1209.0040;%%

%19
\bibitem{LHCHXSWG-www2}
 {\tt https://twiki.cern.ch/twiki/bin/view/LHCPhysics/CrossSections}~.

%20
\bibitem{MSSMHiggsXS}
  E.~Bagnaschi, R.~Harlander, S.~Liebler, H.~Mantler, P.~Slavich and A.~Vicini,
  {\em JHEP} {\bf 1406} (2014) 167
  [arXiv:1404.0327 [hep-ph]].
  %%CITATION = ARXIV:1404.0327;%%

%21
\bibitem{LHCHXSWG-www3} 
 {\tt https://twiki.cern.ch/twiki/bin/view/LHCPhysics/LHCHXSWG}~.

%22
\bibitem{ADLOchargedHiggs} A.~Heister et al.\ [ALEPH Collaboration],
                            {\em Phys.\ Lett.} {\bf B 543} (2002) 1
                            [arXiv:hep-ex/0207054];
                            %%CITATION = PHLTA,B543,1;%%
                            J.~Abdallah et al.\ [DELPHI Collaboration],
                            {\em Eur.\ Phys.\ J.} {\bf C 34} (2004) 399
                            [arXiv:hep-ex/0404012];
                            %%CITATION = EPHJA,C34,399;%%
                            P.~Achard et al.\ [L3 Collaboration],
                            {\em Phys.\ Lett.} {\bf B 575} (2003) 208
                            [arXiv:hep-ex/0309056];
                            %%CITATION = PHLTA,B575,208;%%
                            D.~Horvath {}[OPAL Collaboration],
                            {\em Nucl.\ Phys.} {\bf A 721} (2003) 453.
                            %%CITATION = NUPHA,A721,453;%%

%23
\bibitem{LEPchargedHiggs} G.~Abbiendi et al. [ALEPH and DELPHI
  and L3 and OPAL and LEP Collaborations], 
  {\em Eur.\ Phys.\ J.} {\bf C 73} (2013) 2463
  [arXiv:1301.6065 [hep-ex]].
  %%CITATION = ARXIV:1301.6065;%%

%24
\bibitem{Tevcharged} T.~Aaltonen et al.  [CDF Collaboration],
                     {\em Phys.\ Rev.\ Lett.}  {\bf 103} (2009) 101803
                     [arXiv:0907.1269 [hep-ex]];
                     %%CITATION = ARXIV:0907.1269;%%
                     V.~Abazov et al. [D\O Collaboration],
                     {\em Phys.\ Lett.} {\bf B 682} (2009) 278
                     [arXiv:0908.1811 [hep-ex]];
                     %%CITATION = ARXIV:0908.1811;%%
                     P.~Gutierrez [CDF and D\O Collaborations],
                     PoS CHARGED {\bf 2010} (2010) 004.
                     %%CITATION = POSCI,CHARGED2010,004;%%

%25
\bibitem{LHCcharged} G.~Aad et al. [ATLAS Collaboration],
                     {\em JHEP} {\bf 1206} (2012) 039
                     [arXiv:1204.2760 [hep-ex]];
                     %%CITATION = ARXIV:1204.2760;%%
                     ATLAS Collaboration, 
                     ATLAS-CONF-2013-090; ATLAS-CONF-2014-050;
                     S.~Chatrchyan et al. [CMS Collaboration],
                     {\em JHEP} {\bf 1207} (2012) 143
                     [arXiv:1205.5736 [hep-ex]];
                     %%CITATION = ARXIV:1205.5736;%%
                     CMS Collaboration, 
                     CMS-HIG-13-035; CMS-HIG-14-020.

%26
\bibitem{ILC-TDR}
H.~Baer et al.,
%{\it The International Linear Collider Technical Design Report - Volume 2:
%Physics},
arXiv:1306.6352 [hep-ph].
%%CITATION = ARXIV:1306.6352;%%

%27
\bibitem{MHpLHCILCnewer} A.~Ferrari,
                         talk given at the 
                         {\em C$H^{\mbox{}^\pm}$\hspace{-2.5mm}arged 2006},
                         Uppsala, Sweden, September 2006. 

%28
\bibitem{ppttNNLO} M.~Czakon, P.~Fiedler and A.~Mitov,
  {\em Phys.\ Rev.\ Lett.} {\bf 110} (2013) 252004
  [arXiv:1303.6254 [hep-ph]].
  %%CITATION = ARXIV:1303.6254;%%

%29
\bibitem{Barnett:1987jw} R.~Barnett, H.~Haber and D.~Soper,
                         {\em Nucl.\ Phys.} {\bf B 306} (1988) 697.
                         %%CITATION = NUPHA,B306,697;%%

%30
\bibitem{Harlander:2011aa} R.~Harlander, M.~Kr\"amer and M.~Schumacher,
                           arXiv:1112.3478 [hep-ph].
                           %%CITATION = ARXIV:1112.3478;%%

%31
\bibitem{Flechl:2014wfa}
  M.~Flechl, R.~Klees, M.~Kr\"amer, M.~Spira and M.~Ubiali,
  arXiv:1409.5615 [hep-ph].
  %%CITATION = ARXIV:1409.5615;%%

%32
\bibitem{thdm-lhchxswg-reco}
        R.~Harlander, M.~Muhlleitner, J.~Rathsman, M.~Spira and O.~St{\aa}l,
        arXiv:1312.5571 [hep-ph].
        %%CITATION = ARXIV:1312.5571;%%

%33
\bibitem{hdecay} A.~Djouadi, J.~Kalinowski and M.~Spira,
  {\em Comput.\ Phys.\ Commun.} {\bf 108} (1998) 56
  [arXiv:hep-ph/9704448];
  %%CITATION = HEP-PH/9704448;%%
  A.~Djouadi, M.~M\"uhlleitner and M.~Spira,
  {\em Acta Phys.\ Polon.} {\bf B 38} (2007) 635
  [arXiv:hep-ph/0609292];
  %%CITATION = HEP-PH/0609292;%%
  M.~Spira,
  {\em Fortsch.\ Phys.} {\bf 46} (1998) 203
  [arXiv:hep-ph/9705337].
  %%CITATION = HEP-PH/9705337;%%

%34
\bibitem{2hdmc} D.~Eriksson, J.~Rathsman and O.~St{\aa}l,
  {\em Comput.\ Phys.\ Commun.} {\bf 181} (2010) 189
  [arXiv:0902.0851 [hep-ph]];
  %%CITATION = ARXIV:0902.0851;%%
  {\em Comput.\ Phys.\ Commun.} {\bf 181} (2010) 833.
  %%CITATION = CPHCB,181,833;%%

%35
\bibitem{feynhiggs} S.~Heinemeyer, W.~Hollik and G.~Weiglein,
                    {\em Comput. Phys. Commun.} {\bf 124} (2000) 76, 
                    [arXiv:hep-ph/9812320];
                    %%CITATION = HEP-PH 9812320;%%
           T.~Hahn, S.~Heinemeyer, W.~Hollik, H.~Rzehak and G.~Weiglein,
           {\em Comput.\ Phys.\ Commun.} {\bf 180} (2009) 1426;
           %%CITATION = CPHCB,180,1426;%%
           see: {\tt www.feynhiggs.de} .

%36
\bibitem{mhiggslong} S.~Heinemeyer, W.~Hollik and G.~Weiglein,
                     {\em Eur. Phys. J.} {\bf C 9} (1999) 343
                     [arXiv:hep-ph/9812472].
                     %%CITATION = HEP-PH 9812472;%%

%37
\bibitem{mhiggsAEC} G.~Degrassi, S.~Heinemeyer, W.~Hollik,
                    P.~Slavich and G.~Weiglein, 
                    {\em Eur. Phys. J.} {\bf C 28} (2003) 133
                    [arXiv:hep-ph/0212020].
                    %%CITATION = HEP-PH 0212020;%%

%38
\bibitem{mhcMSSMlong} M.~Frank, T.~Hahn, S.~Heinemeyer, W.~Hollik,  
                      H.~Rzehak and G.~Weiglein,
                      {\em JHEP} {\bf 0702} (2007) 047
                      [arXiv:hep-ph/0611326].
                      %%CITATION = HEP-PH 0611326;%%

%39
\bibitem{mhcMSSM2L} S.~Heinemeyer, W.~Hollik, H.~Rzehak and G.~Weiglein,
                    {\em Phys. Lett.} {\bf B 652} (2007) 300
                    [arXiv:0705.0746 [hep-ph]].
                    %%CITATION = ARXIV:0705.0746;%%

%40
\bibitem{Mh-logresum} T.~Hahn, S.~Heinemeyer, W.~Hollik, H.~Rzehak and
                      G.~Weiglein,  
{\em Phys. Rev. Lett.} {\bf 112} (2014) 141801
[arXiv:1312.4937 [hep-ph]].
%%CITATION = ARXIV:1312.4937;%%

%41
\bibitem{cpsh} J.~Lee, A.~Pilaftsis et al.,
               {\em Comput. Phys. Commun.} {\bf 156} (2004) 283
               [arXiv:hep-ph/0307377];
               %%CITATION = HEP-PH 0307377;%%
               J.~Lee, M.~Carena, J.~Ellis, A.~Pilaftsis and C.~Wagner,
               {\em Comput.\ Phys.\ Commun.} {\bf 180} (2009) 312
               [arXiv:0712.2360 [hep-ph]];
               %%CITATION = ARXIV:0712.2360;%%
               arXiv:1208.2212 [hep-ph].
               %%CITATION = ARXIV:1208.2212;%%

%42
\bibitem{deltamb1} R.~Hempfling,
                   {\em Phys. Rev.} {\bf D 49} (1994) 6168;
                   %%CITATION = PHRVA,D49,6168;%%
                   L.~Hall, R.~Rattazzi and U.~Sarid,
                   {\em Phys. Rev.} {\bf D 50} (1994) 7048,
                   hep-ph/9306309;
                   %%CITATION = HEP-PH 9306309;%%
                   M.~Carena, M.~Olechowski, S.~Pokorski and C.~Wagner,
                   {\em Nucl. Phys.} {\bf B 426} (1994) 269,
                   hep-ph/9402253.
                   %%CITATION = HEP-PH 9402253;%%

%43
\bibitem{deltamb2} M.~Carena, D.~Garcia, U.~Nierste and C.~Wagner,
                   {\em Nucl. Phys.} {\bf B 577} (2000) 577,
                   hep-ph/9912516.
                   %%CITATION = HEP-PH 9912516;%%

%44
\bibitem{db2l} D.~Noth and M.~Spira,
               {\em Phys.\ Rev.\ Lett.} {\bf 101} (2008)  181801
               [arXiv:0808.0087 [hep-ph]].

%45
\bibitem{benchmark2} M.~Carena, S.~Heinemeyer, C.~Wagner and G.~Weiglein, 
                     {\em Eur. Phys. J.} {\bf C 26} (2003) 601
                     [arXiv:hep-ph/0202167].
                     %%CITATION = HEP-PH 0202167;%%

%46
\bibitem{BR} A.~Denner, S.~Heinemeyer, I.~Puljak, D.~Rebuzzi and M.~Spira,
             {\em Eur.\ Phys.\ J.} {\bf C 71} (2011) 1753
             [arXiv:1107.5909 [hep-ph]].
             %%CITATION = ARXIV:1107.5909;%%

%47
\bibitem{benchmark4}
  M.~Carena, S.~Heinemeyer, O.~St{\aa}l, C.~Wagner and G.~Weiglein,
  {\em Eur.\  Phys.\  J.} {\bf C 73} (2013) 2552
  [arXiv:1302.7033 [hep-ph]].
  %%CITATION = ARXIV:1302.7033;%%